\documentclass[a4paper]{article}
\usepackage{epsfig,amsmath,amsfonts,amssymb,setspace,multirow,textcomp}
\makeatletter
\renewcommand{\@oddhead}{\textit{Advances in Astronomy and Space Physics} \hfil}
\renewcommand{\@evenfoot}{\hfil \thepage \hfil}
\renewcommand{\@oddfoot}{\hfil \thepage \hfil}
\makeatother

\renewenvironment{thebibliography}[1]{\begin{oldthebibliography}{#1}\setlength{\parskip}{0ex}\setlength{\itemsep}{0ex}}{\end{oldthebibliography}}

\begin{document}
\fontsize{11}{11}
\selectfont 
\title{Constraining the intrinsic energy of GRB events }
\author{\textsl{E.\,D.~Lopez$^{1,2}$}}
\date{\vspace*{-6ex}}
\maketitle
\begin{center} {\small $^{1}$ Quito Astronomical Observatory of National Polytechnic School, Interior del parque La Alameda,\\ Av. Gran Colombia s/n, Quito, Ecuador.\\
$^{2}$Space Telescope Science Institute, 3700 San Martin Drive, Baltimore, MD 21218, USA.\\
{\tt ericsson.lopez@epn.edu.ec}}
\end{center}

\begin{abstract}
Considering a GRB event as a relativistic ejecta where the relativistic moving makes radiation become anisotropic, we are able to show that the required intrinsic energy associated with these events is significantly smaller than those values commonly presented in literature for an isotropic distribution of emitted energy. Our results show energy values around $10^{44}$ ergs for Lorentz $\Gamma$  factor $\sim 10$ and around $10^{38}$ ergs for $\Gamma \sim 300$, values which are more compatible with energies involved in AGN events rather than those related to the formation of stellar black holes and hypernovas.\\[1ex]
\end{abstract}

\section*{\sc introduction}
\indent \indent 
The first confident detection of a gamma-ray burst (GRB) was made by the Vela IVa satellite system in 1967 (see e.g. \cite{bo96}) , although the first paper about the discovery of GRBs was published by \cite{kl73} in 1973. The VELA sensors were able to detect primary X-ray signatures and complementary neutron and gamma-ray radiation from exoatmospheric clandestine nuclear detonations. Inside the big question about the localization of the first GRB events, the distribution of the burst seemed to imply an isotropic distribution of sources and did not seem to favor the Galactic origin of the bursts. This uncertainty over the localization and nature of GRBs remained until multifrequency observations become available (\cite{kl12}). This was a pioneering phase of confusion between 1967 until nearly 1991. 

With the advent of the Burst and Transient Source Experiment (BATSE) of the Compton Gamma Ray Observatory (CGRO), instrument which operates continuously since 1991 to 2000 year, a final decision was taken on the debates of the Galactic Vs. extragalactic origin of GRB events. On the base of a large homogeneous data collection of GRBs the Cosmological hypothesis won this dispute with models in which the isotropy was naturally explained. The BATSE discovery of the isotropic distribution of the faintest bursts reinforced this fundamental conclusion (see e.g. \cite{ma92}). 

The primary objections to the cosmological hypothesis were the compactness problem and the required high GRB luminosities of the order of $10^{51} erg$ or even more in few seconds. Moreover, the observed short variability time scales ($\Delta t \ll 1 sec $) also required a convincing physical explanation.

For cosmological GRBs an instantaneous energy of $10^{51}$ ergs is implied from the observed flux of about $10^{-7}$  ${ergs\over {cm^2 sec}}$, for this value both the absorptive and scattering optical depths are very large and it is extremely difficult to understand how photons about the   pair creation threshold escape from the emitting region, located close to the compact source. This problem was resolved by invoking relativistic outflows of a pair plasma, in the so-called fireball/blast wave model (great concentration of photons confined in a small space) proposed by \cite{go86} and \cite{pa86}, where the high opacity diminishes in the violent relativistic expansion and the radiation is observed. The observed non-thermal gamma-ray spectra was explained with the conversion of kinetic energy to gamma radiation at external (and latter internal) shocks (\cite{me93}).     

The main contribution of BATSE experiment was to provide good statistic data to conclude that the sources of isotropic distributed GRBs were at cosmological distances. Besides that, not less important contribution was that the temporal and spectral properties of the large number of GRBs were observed in great detail and in particular the prompt phase. 

The COMPTON experiments open the way, for subsequences satellite missions as BeppoSAX, Hete-2, Integral, Swift and other, to the GRB cosmological era.

In the current work, we briefly describe an alternative GRB model, proposed recently, that takes in consideration the relativistic bulk motion of the plasma and the jet geometry. This work is attempting to provide a further contribution to understanding the physics and origin of GRBs events.

\section*{\sc relativistic motion}

\indent \indent Models with relativistic motion are adopted to resolve the compactness problem (see e.g., \cite{go86, pa86, kr91}). Usually, the extremely large opacity is reduced by including into the calculations the relativistic expansion of the source which is moving toward us. However, this alternative introduces new difficulties,  for instance,  a $\Gamma$ Lorentz factor  $\sim 100$ is required to guarantee the transparence of the medium ($\tau < 1$ for energies $E'< 10^{38}$ ergs). Thus, we must to justify how the necessary energy to produce extremely relativistic motion of the GRB sources is provided. Consequently, a serious problem with the energetic of these extragalactic objects has been arisen.

In this work, we are following the same assumption, considering the large opacity problem as a relativistic illusion provoked by the bulk relativistic motion of the emitting $\gamma$ ray plasma. In this context, the main parameters, which characterize the physical conditions of the emitting material, must be reduced or boosted by a suitable potency of the Doppler Lorentz factor  $\delta = {1/ {\Gamma (1-\beta \mu)}}$, where $\Gamma = {1\over{\sqrt{1- \beta ^2}}}$, $\beta = v/c$, $ \mu= \cos \theta$, $ v$ is the flow velocity,  $\theta$ is the angle formed between the velocity direction and the line of sight  and $c$ is the speed of the light. The main idea behind the present contribution is to consider that due the bulk motion of the emitting plasma, the radiation received by the Earth's observer is not more isotropic, therefore a suitable expression for the transformation of the flux density ($F_\nu$) must be derived. 

\section*{\sc anisotropic fireball model}
\indent \indent

Considering the global motion of the emitting plasma, the optical depths for both absorptive and scattering processes are reduced in the lab frame (for the observer) by the Doppler factor $\Gamma$ as $\tau = \delta ~ \tau' \sim {1\over{2 \Gamma^2}}~ \tau'$, where  $\tau$ is the optical depth at the observer frame and $\tau'$ at the source one, $\delta = {1\over {\Gamma (1 -\beta \mu)}}$. This result is in concordance with the expression given by \cite{kr91} (see also Piran \& Shemi 1993). On the other hand, the relativistic transformation of specific intensity is derived in an elementary way by transforming the photon number densities and the energy by the relativistic aberration of the angle $\theta$ (angle between the line of sight and the flux direction).  Therefore, for a moving source, the observed monochromatic flux density $F_\nu$ is related to the flux density in the comoving frame $F'_\nu$  by the expression $F_\nu = ({\nu \over{\nu'}})^3  F'_\nu$, where $\nu$ is the observed frequency and $\nu'$ is the frequency related to the comoving frame. Consequently, the total fluxes are connected  in both frames by   $F = (\delta^4)  F'$, where $ \delta$ is the Doppler Lorentz factor. Integrating the total flux on a closed surface, we obtain the luminosity $L$ that  gives us the power of energy released by the source. In most of the proposed models the isotropic radiation cannot provide the necessary energy for the appearance of a cosmological GRB (\cite{bi06}). Therefore, we must considerer an anisotropic distribution for the observed radiation (\cite{lo12}). 

In order to address the energetic problem of GRBs, their origin and variability, we have proposed an alternative model, where at the rest frame of the source the radiation is released isotropically, so the flux $F'$ does not depend on the angular coordinates (i.e., $F'= F'(r)$).  However, it is assumed that the radiation is emitted from a source which is moving highly relativistically (fireball/relativistic blast wave model). Then, in concordance with that exposed by others authors (see e.g., \cite{ghi99,lo04}), the observed radiation must be affected by the boosting Lorentz factor $\delta$. Consequently, we would to detect in the laboratory frame a flux $F$ enhanced by the  $\delta$ factor ($ F = \delta^4 ~ F'$), and at the observed frame we should expect an anisotropic flux which is dependent on the propagation direction of radiation ($F = F(r,\theta, \phi)$).

The total power emitted by the source and detected by the Earth's observer (luminosity L) should be computed integrating over a closed surface enclosing the source and how it was mentioned above in this integration we should incorporate the axial dependence of flux due to the relativistic beaming. Consequently, the integrated power emitted by the source at the observed frame yields (\cite{lo12}): 
$L =  {4\pi r^2 \over 3} ( 4 \Gamma ^2 -1) [\Gamma - \sqrt{\Gamma ^2 - 1 } ]^4  F(0^\circ)$, where $F(0^\circ)$ is the observed flux when the jet orientation coincides with the line of sigh ( $\theta = 0^\circ$ ). Now, considering that typically a GRB is lasting 1 second, the total intrinsic energy of the source can be deduced. The final expression that we have derived is (\cite{lo12}):

$$ E'= {4\pi r^2 \over 3} ( 1 - {1\over {4 \Gamma ^2}}) [\Gamma - \sqrt{\Gamma ^2 - 1 } ]^4  F(0^\circ)~ t.\eqno(1) $$

The above expresion provides the intrinsic energy released in a GRB event according the anisotropic model where the relativistic boosting and the jet geometry have been taken into account in the calculations.

In the table shown below, we present the results obtained for the intrinsic energy of a GRB event for sources located at $1~ GPc\sim 3\times 10^{27} cm$ and at $3 ~GPc \sim 1 \times 10^{28} cm $, for two values of observed fluxe: $F (0^\circ) \sim 10^{-6} {ergs\over{cm^2~ sec}}$ and $F (0^\circ) \sim 10^{-7} {ergs\over{cm^2~ sec}}$ for several values of the $\Gamma$ Doppler factor.

\begin{table}[h]
\centering
 \caption{Intrinsic energy released in a GRB event (\cite{lo12}).}\label{tab1}
 \begin{tabular}{cccc}
  \hline
  Cosmological Distance & Observed Flux  & Lorentz Factor & Intrinsic Energy 
\\ r ~(cm)& $F(0^{\circ})$ $(ergs/cm^2~ s)$ &$\Gamma $ &E'$(ergs)$\\
  \hline
 $3 \times 10^{27}$  & $1\times 10^{-6}$ & $ 10$ & $2.37 \times 10^{44}$ \\
 &  & $100$  & $2.36 \times 10^{40}$ \\
 &  & $150$  & $4.65\times 10^{39}$ \\
 &  & $300$  & $2.9\times 10 ^{38}$\\
\hline
$3 \times 10^{27}$  & $1\times  10^{-7}$ & $ 10$ & $2.3 \times 10^{43}$ \\
  &  & $100$  & $2.36 \times 10^{39}$ \\
  &  & $150$  & $4.65 \times 10^{38}$ \\
  &  & $300$  & $2.9\times 10 ^{37}$\\
\hline
$1 \times 10^{28}$  & $1\times 10^{-6}$ & $ 10$ & $2.64 \times 10^{45}$ \\
  &  & $100$ & $2.62 \times 10^{41}$ \\
  &  & $150$  & $5.2 \times 10^{40}$ \\
  &  & $300$  & $3.23\times 10 ^{39}$\\
\hline
$1 \times 10^{28}$  & $1\times 10^{-7}$ & $ 10$ & $ 2.64\times 10^{44}$ \\
  &  & $100$ & $2.62 \times 10^{40}$ \\
  &  & $150$  & $5.2 \times 10^{39}$ \\
  &  & $300$  & $3.23\times 10 ^{38}$\\
\hline

\end{tabular}
\end{table}

\section*{\sc discussion and conclusions}
\indent \indent 
Several authors have made evident the fact that the energy released in a gamma-ray event could be overestimated if the emission is considered isotropic (see e.g., \cite{fra01,ghi99}). In these models the energy involved is extremely large leading to powers of about $10^{52}~{ergs\over{sec}}$- $10^{54} ~{ergs\over{sec}}$ in a single event. However, the observations carried out on these explosive events suggest us that anisotropic models are also a good alternative and maybe a more realistic suggestion.

The radiation in the intrinsic frame can be isotropic, however the observed radiation at the laboratory frame, due to the relativistic beaming, is confined in a very small opening angle and the flux becomes anisotropic. This point of view does not require large values for the Lorentz gamma factor and the intrinsic energy associated with a gamma-ray event is greatly reduced with the assumption of the anisotropic character of the observed radiation (see above results). Following  this  approach, we derive that for relatively small values of $\gamma \sim 10$ the true energy delivered in a gamma-ray event is about $1\times10^{44}~ ergs$, whereas for  $\gamma \sim 100$ the energy is about $1\times10^{40}~ ergs$ and for $\gamma \sim 300$ around $1\times10^{38}~ ergs$, values which are significantly smaller than those commonly presented in the literature.

Our results show that the intrinsic energy released in a gamma-ray event could be smaller than the typical electromagnetic and kinetic energy produced in ordinary supernovae. Therefore, the GRB events could be not necessarily associated with the formation of stellar black holes or hypernovae as it was previously suggested. The values present in this paper are more compatible with the energies  involved in AGN events, where a fraction of a solar mass per year can be accelerated to $\gamma \sim 10$, leading to powers of $\sim 10^{46} ~ {ergs\over {sec}}$.

\section*{\sc acknowledgement}
\indent \indent The author wants to express his gratitude to Space Telescope Science Institute (STScI) for the huge support and necessary facilities provided for the preparation of this work during his visiting scholar program. E.L. was supported by the National Secretary of Higher Education, Science, Technology and Innovation of Ecuador (Senescyt, fellowship 2011).

\end{document}